\documentclass[reprint,superscriptaddress,floatfix]{revtex4-1}
\usepackage{amsmath}
\usepackage{amssymb}
\usepackage{color}
\usepackage{siunitx}
\usepackage{graphicx}
\usepackage{dcolumn}
\usepackage{bm}
\usepackage{hyperref}

\usepackage[T1]{fontenc}
\usepackage{lmodern} 

\usepackage{multirow}
\usepackage{booktabs}
\usepackage[table,xcdraw]{xcolor}

\begin{document}
\title{\emph{Ab initio} Theory of Eliminating Surface Oxides of Superconductors with Noble-Metal Encapsulation}

\author{Crist\'obal M\'endez}
\affiliation{School of Applied and Engineering Physics, Cornell University, Ithaca, New York 14853, USA}
\author{Nathan Sitaraman}
\affiliation{Department of Physics, Cornell University, Ithaca, New York 14853, USA}
\author{Matthias Liepe}
\author{Tom\'as A. Arias}
\affiliation{Department of Physics, Cornell University, Ithaca, New York 14853, USA}

\date{\today}

\begin{abstract}
Nanometer-scale surface chemistry limits the performance of SRF cavities and quantum circuits. We present an \emph{ab initio} framework connecting DFT interfacial energetics with strong-coupling Eliashberg theory for capped Nb and Ta surfaces. This approach identifies Au and Au-based alloys (AuPd, AuPt) as effective passivation layers. Our model further predicts that combining a noble-metal capping layer with an appropriate wetting/adhesion layer (WAL) yields far more robust adhesion than a capping layer alone under realistic conditions, enabling thinner caps, and thereby addressing a central challenge in superconducting surface passivation.
\end{abstract}

\maketitle
Niobium (Nb) and tantalum (Ta) are cornerstones of today’s highest-performance superconducting resonator technologies.
Nb remains the canonical material for superconducting radio-frequency (SRF) cavities, where the quality factor $Q$ is limited by surface resistance arising from processes within only a few tens of nanometers of the surface~\cite{yu2022oxidation,padamsee201750}.
Ta, in turn, has emerged as a premier substrate for superconducting qubits, delivering record $Q$ relative to Nb for low-field 2D resonators~\cite{de2021materials,bal2024systematic,place2021new}.
For both metals, a strong affinity for oxygen and hydrogen drives the formation of pentoxides at the surface and interstitial hydrogen (H)~\cite{antoine2012materials,yu2022oxidation,ramirez2010amorphous}.
These defects seed two-level systems (TLS) in qubits and contribute to residual loss in SRF cavities, making monolayer-scale surface chemistry a primary determinant of device performance~\cite{bal2024systematic,valente2016superconducting}.

Metallic capping layers provide an industrially compatible route for stabilizing Nb and Ta surfaces during processing and operation, thereby enabling air-stable devices.
Reported strategies, however, span disparate approaches:  
(i) noble-metal films that chemically passivate~\cite{huger2005alloying,huger2005subsurface,murthy2022developing,oseroff2023measurements,karuppannan2024improved},  but these must be kept ultrathin to avoid disturbing superconductivity;  
(ii) superconducting Ta caps on Nb, shown experimentally to enhance Nb-based qubit coherence by replacing Nb’s native oxide with a less lossy Ta oxide~\cite{bal2024systematic}; and  
(iii) more recently, AuPd encapsulation on Ta circuits, which reduces the capping thickness needed to suppress oxide formation compared with pure Au encapsulation~\cite{chang2025eliminating}.  

Outcomes vary because three coupled constraints must be met simultaneously. First and foremost, an ultrathin cap must wet and adhere across surface imperfections under realistic environmental conditions, which is a central unsolved challenge for ultrathin caps at present~\cite{paradkar2025superconducting,seddon2024thin,chang2025eliminating,wang2025epitaxial}. Such a cap also must chemically passivate against O/N/H species that generate TLSs. Finally, the cap must respect a superconducting thickness budget, since normal-metal caps depress the substrate’s \(T_c\) via the proximity effect~\cite{schachinger1984critical,martinis2000calculation,ummarino2023strong}.  
Experiments therefore show that both the material and the thickness of the cap are critical~\cite{bal2024systematic,chang2025eliminating}.

\begin{figure}[htbp]
  \centering
  \includegraphics[width=0.7\columnwidth]{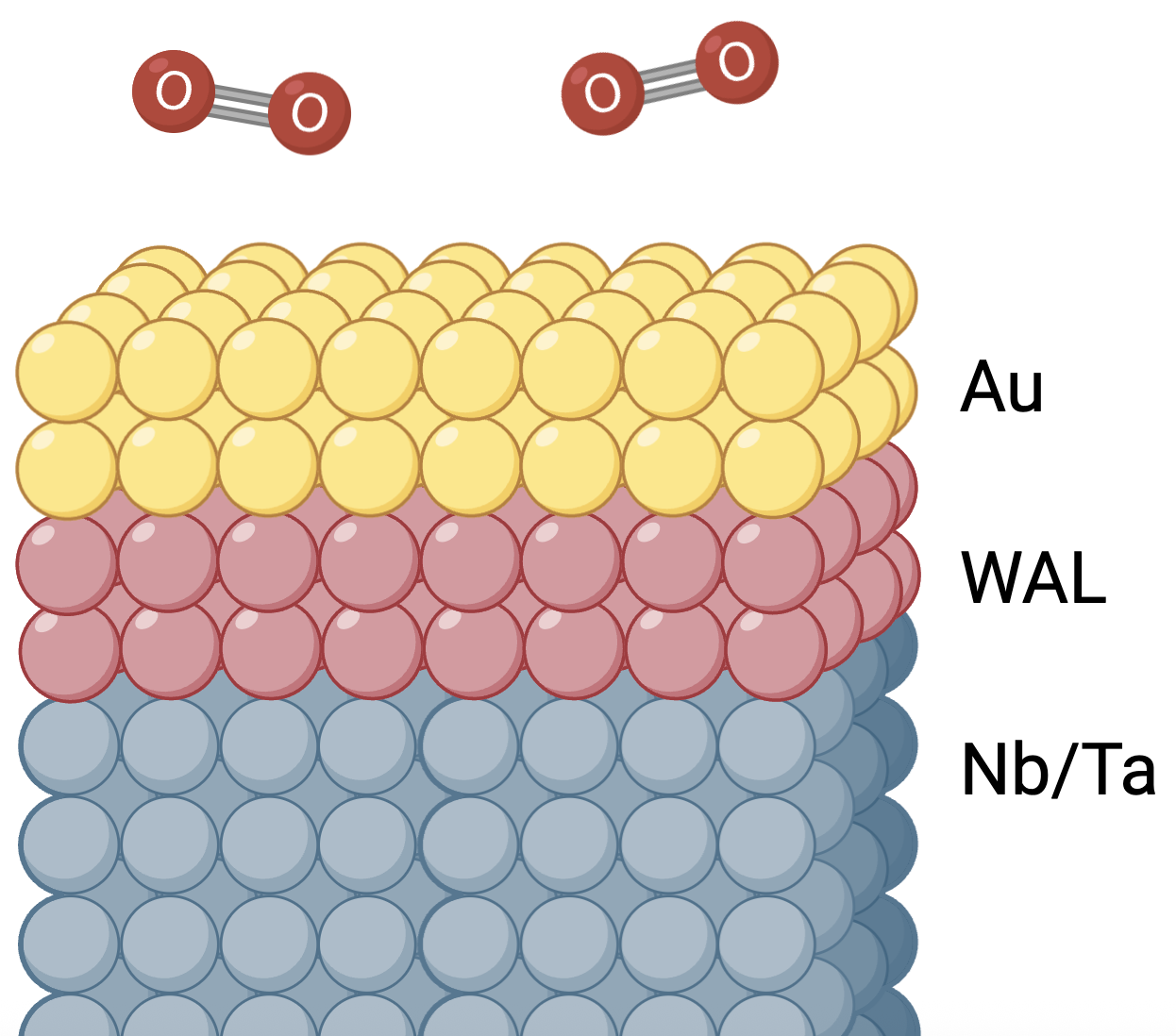}
  \caption{Illustration of Au capping layer, wetting/adhesion underlayer (WAL), and Nb/Ta substrate.}
  \label{fig:wal_diagram}
\end{figure}

Here we develop a unified, first-principles framework that treats the chemically related metals Nb and Ta on equal footing.  
Our approach combines density-functional calculations of both interfacial energetics and impurity adsorption with Eliashberg theory for proximity-coupled bilayers~\cite{ummarino2023strong,kubo2024applicability}. Within this framework, we identify two potential roles in any practical stack of ultrathin capping layers:  
\emph{(i) a passivation layer} that terminates the surface and weakly binds O/N/H;  
\emph{(ii) a wetting/adhesion underlayer (WAL)}---introduced here to decouple adhesion from passivation---placed directly between the substrate and the passivation layer, as a way to maximize adhesion while keeping the passivation layer as thin as possible (Fig.~\ref{fig:wal_diagram}).We also briefly discuss the possibility of idealized sacrificial getter layers, opposite in intent to passivation, whose role is to capture oxygen by maximizing oxygen affinity.

We then determine which metals robustly fulfill each role on Nb(110) and Ta(110), quantify the allowable normal-metal thicknesses without suppressing \(T_c\), and derive compact design rules that account for recent experimental findings---notably, enhanced coherence in Nb qubits capped with Ta and Au~\cite{bal2024systematic}, and the success of Ta qubits encapsulated with AuPd alloys over Au~\cite{chang2025eliminating}. Guided by these rules, we propose new candidate material stacks for SRF cavities and superconducting qubits that mitigate surface loss and extend coherence times. The following sections present the underlying calculations and resulting guidelines.

\par\medskip

\emph{Computational Details.---}
All \textit{ab initio} DFT calculations were performed using our open-source plane-wave software JDFTx \cite{payne1992iterative,sundararaman2017jdftx}, employing ultrasoft pseudopotentials \cite{garrity2014pseudopotentials} and the PBEsol functional \cite{perdew2008restoring}. A plane-wave cutoff of 20 Hartree was used for the electronic wave functions, and 100 Hartree for the electron density. A Fermi smearing corresponding to an effective temperature of 4~mHa was applied for the Kohn–Sham orbital occupancies. Bulk calculations utilized $3\times3\times3$ supercells of the original cubic cells and used a \textit{k}-point density corresponding to  $12\times12\times12$ in the primitive cell, while surface energy calculations employed a 12-layer slab supercell with truncated Coulomb potentials in the perpendicular direction to improve accuracy \cite{sundararaman2013regularization}.
For our equiatomic AB alloy caps, we constructed ultrathin films composed of ordered monolayers (ML) on Nb/Ta(110), exploring both A/B/A/B and inverted B/A/B/A cap-layer sequences on the substrate surface. The most stable configurations feature a Pd/Pt-rich interface layer and an Au- or Ag-rich surface termination.

\par\medskip
\emph{Passivation layers.---}
A first metric for selecting passivation materials is their bulk affinity for oxygen, nitrogen, and hydrogen, which should be as weak as possible.
Accordingly, we surveyed 25 elemental metals chosen to cover the 3d–5d transition series and a representative selection from groups 2–14. We deliberately restricted the screening to elemental metals to obtain a systematic map of periodic trends in adsorption and wetting that can be traced back to basic electronic structure, rather than compound-specific chemistry. We also included metals that are already used as caps or adhesion layers in Nb/Ta devices, so the survey reflects fabrication-relevant choices. The resulting periodic trends and design rules provide a baseline for future studies of inorganic and organic capping materials.

For each candidate, we compute the interstitial formation energy
\[
E_{\text{inter}}^{X}=E_{\text{bulk+}X@{\rm int}}-E_{\text{bulk}}-\tfrac12E_{X_2},
\qquad X=\mathrm{O,N,H},
\]
taking the lowest energy among high-symmetry interstitial sites, and using the diatomic molecule as the reference chemical potential. Positive values imply that the impurity will tend to be rejected from the material.
 
Table~\ref{tab:interstitials} summarizes our results for the 25 elemental metals we surveyed. Nb and Ta readily incorporate interstitial O, N, and H (negative $E_{\mathrm{inter}}$), whereas several relatively late, refractory metals---W, Ru, Ag, Au, Pt, Rh, Ir, and Pd---exhibit positive formation energies for O and N, indicating unfavorable bulk incorporation and making them attractive candidates for vacuum-facing layers.

These trends broadly track simple electronic considerations: early transition metals (groups 3--5; e.g., Hf, Zr, Ti, Y, Sc) have higher $d$-band centers and lower electronegativities, which strengthen $p\!-\!d$ bonding with O/N and often stabilize hydrides, yielding strongly negative $E_{\mathrm{inter}}$ and making them effective getters. By contrast, the more filled $d$ bands of late transition metals weaken hybridization with O/N (and in many cases H), pushing $E_{\mathrm{inter}}$ upward and favoring impurity rejection. Among potential getters, we focus on Zr, given its established benefits in Nb-based SRF cavities~\cite{sitaraman2023enhanced}, while noting that Hf, Ti, Y, and Sc show similarly low $E_{\mathrm{inter}}$.

Turning to wetting, because the bcc(110) surface has the lowest energy for \emph{both} substrates, all surface and interface calculations use this orientation. We first compute the (110) surface energy of the Nb or Ta substrate, $\gamma_\mathrm{sub(110)}$. Next, for each capping metal (or alloy) we evaluate the interface energy \(\gamma_{\mathrm{int}}\) between the metal and Nb(110) or Ta(110) and also the epitaxial free surface energy \(\gamma_{\mathrm{cap}}\) of the same overlayer constrained to the substrate lattice.  
Wetting is gauged with the spreading parameter
$S=\gamma_{\mathrm{int}}+\gamma_{\mathrm{cap}}-\gamma_{\text{sub}(110)}$; wetting occurs when \(S<0\).

\begin{figure}[htbp]
  \centering
  \includegraphics[width=\columnwidth]{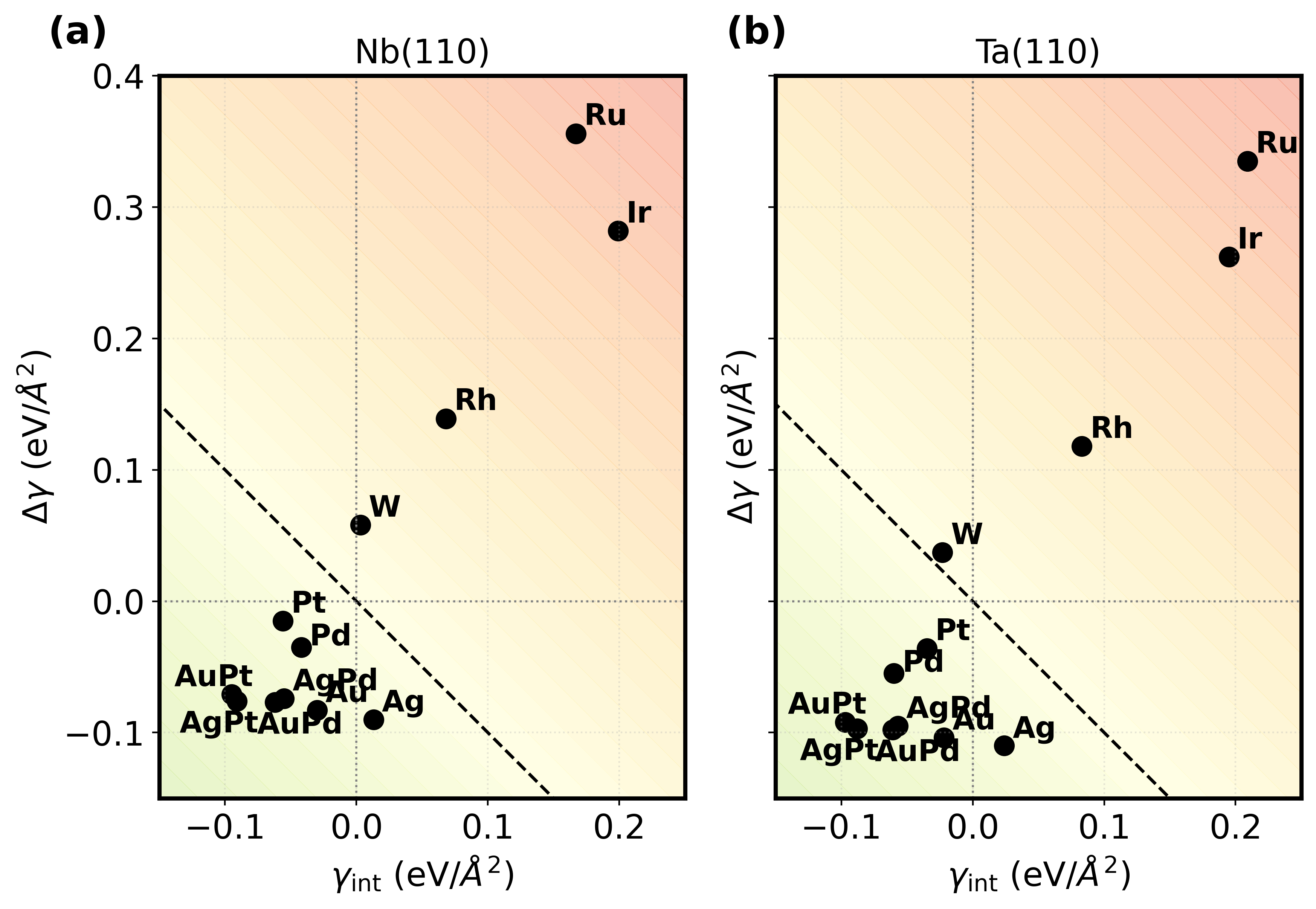}
  \caption{Interface map for candidate caps on (a) Nb(110) and (b) Ta(110).  
           Axes: interface energy \(\gamma_{\mathrm{int}}\) vs surface-energy contrast \(\Delta\gamma=\gamma_{\mathrm{cap}}-\gamma_{\text{sub}(110)}\).
           Dashed line: \(S=0\) wetting boundary; shaded gradient: wetting (green) to islanding (red).}
  \label{fig:islandMaps}
\end{figure}

Figures~\ref{fig:islandMaps}(a,b) plot \(\gamma_{\mathrm{int}}\) against \(\Delta\gamma=\gamma_{\mathrm{cap}}-\gamma_{\text{sub}(110)}\) for different caps on Nb and Ta, respectively.  
On ideal (110) terraces, we find that for both Nb and Ta, the metals Au, Ag, Pd, and Pt fall squarely inside the wetting quadrant for ultrathin films. Physically, this favorable wetting can be understood in terms of $d$-band filling at early/late transition-metal interfaces: when the combined interfacial $d$-electron manifold is closer to half filling on average, interfacial bonding is strengthened, which lowers the interface energy $\gamma_{\mathrm{int}}$ and thereby promotes wetting.

Equiatomic binary alloys (in 50:50 proportion) that pair Au or Ag with Pd or Pt are driven deeper into the wetting region in our DFT calculations because, after full relaxation, they self-segregate into a Pd/Pt-rich interfacial monolayer (lowering $\gamma_{\mathrm{int}}$) and an Au/Ag-rich surface monolayer (lowering $\gamma_{\mathrm{cap}}$). Finally, for Ta we note that its larger surface energy shifts \(\Delta\gamma\) downward and improves wetting, but not enough to stabilize any additional metals in our study, leaving the same four elemental metals and their Au-rich alloys on the wetting side.

Metals that both exhibit poor bulk incorporation of O and wet Nb(110) and Ta(110)—notably Au, Pd, Pt, and Au-rich Pd/Pt alloys—emerge as prime outer-layer candidates, while their alloys (AuPd, AuPt, AgPd, AgPt) exhibit even better wetting behavior, consistent with the
experimental finding that thinner alloy caps suffice to suppress oxide formation on Ta superconducting
chips~\cite{chang2025eliminating}.

\begin{table}
  \caption{Interstitial formation energies (in eV)  \(E_{\text{inter}}^{X}\) (\(X=\mathrm{O,N,H}\)) for candidate metals.}
  \label{tab:interstitials}
  \begin{ruledtabular}
    \begin{tabular}{l d d d}
      Element &
      \multicolumn{1}{c}{$E_{\mathrm{inter}}^{\mathrm{O}}$ (eV)} &
      \multicolumn{1}{c}{$E_{\mathrm{inter}}^{\mathrm{N}}$ (eV)} &
      \multicolumn{1}{c}{$E_{\mathrm{inter}}^{\mathrm{H}}$ (eV)} \\
      \hline
      Ir  &  1.80 &  2.75 &  0.78 \\
      Rh  &  1.44 &  1.44 &  0.05 \\
      Pt  &  1.18 &  1.99 &  0.45 \\
      Au  &  0.85 &  3.70 &  0.69 \\
      Pd  &  0.36 &  1.46 & -0.27 \\
      Ag  &  0.20 &  3.35 &  0.47 \\
      Ru  &  0.11 &  1.19 &  0.05 \\
      W   &  0.07 &  1.07 &  0.75 \\
      Cu  & -0.20 &  1.94 &  0.25 \\
      Mo  & -0.82 &  0.30 &  0.41 \\
      Cr  & -1.36 & -0.25 &  0.44 \\
      Pb  & -1.97 &  1.78 &  0.79 \\
      Tc  & -2.56 & -1.33 & -0.40 \\
      Cd  & -2.94 &  1.41 &  0.87 \\
      Al  & -4.14 & -1.09 &  0.60 \\
      Ta  & -4.32 & -2.74 & -0.47 \\
      Nb  & -4.41 & -2.62 & -0.51 \\
      V   & -4.99 & -3.12 & -0.49 \\
      Ba  & -5.05 & -0.28 & -0.62 \\
      Mg  & -5.95 & -2.07 & -0.11 \\
      Hf  & -7.34 & -4.68 & -0.55 \\
      Zr  & -7.37 & -4.54 & -0.72 \\
      Ti  & -7.42 & -4.63 & -0.74 \\
      Y   & -7.49 & -3.82 & -0.99 \\
      Sc  & -7.73 & -4.25 & -1.05 \\
    \end{tabular}
  \end{ruledtabular}
\end{table}

\paragraph*{Coherent thickness and strain.}
Optimal wetting and passivation require coherently strained capping layers. Achieving this across differing crystal structures (e.g., fcc films on bcc substrates) and lattice constants can demand substantial elastic accommodation, such as Bain-type tetragonal deformations of the fcc overlayer. Accordingly, we define the epitaxial elastic energy (per unit area per monolayer) \(\varepsilon_{\mathrm{strain}} = E_{\text{bcc}}-E_{\text{bulk}}\), where the computation of $E_{\text{bcc}}$ fixes the surface components of the strain to the substrate and allows full relaxation of the atomic coordinates (including the Bain transformation for fcc overlayers) and the vertical component of the strain. The calculated values for $\varepsilon_{\mathrm{strain}}$, along with the lattice mismatch of the relaxed bcc phases of each material with respect to $a_{\text{Nb}}\sim a_{\text{Ta}} \sim 3.30$~\AA, are shown in Table~\ref{tab:systems}. A simple energetic upper bound for the number of coherent monolayers is $N_{\mathrm{ML}}\;\approx\;\frac{|S|a_{\parallel}}{\varepsilon_{\mathrm{strain}}}$,
where \(S\) is the spreading parameter and \(a_{\parallel}\) is the area per surface atom on the (110) facet. Table~\ref{tab:systems} lists the resulting \(N_{\mathrm{ML}}\) for materials with \(S<0\) using DFT-calculated parameters. Au and Ag can remain coherent for several tens of monolayers on either substrate, whereas Pd and Pt are limited to only a few, with alloys falling in between. Beyond this thickness, we can expect large numbers of misfit dislocations and other relaxation defects, providing pathways for impurity incorporation. Because the energetic margin is quickly exhausted with increasing thickness, full coverage must be secured at the few-monolayer level rather than by simply thickening the film.

\squeezetable 
\begin{table}
  \caption{Estimated maximum coherent thickness $N_{\mathrm{ML}}$ (in monolayers) for candidate caps on Nb(110) and Ta(110), using the epitaxial strain energy $\varepsilon_{\text{strain}}$ and spreading parameter $S$. For elemental and alloy caps, we also report the bcc lattice mismatch relative to Nb/Ta (using $a_{\text{Nb}}\approx a_{\text{Ta}}\approx 3.30$~\AA). The lower block shows the effect of a Cu wetting/adhesion underlayer (WAL) and surface microstructure.}
  \label{tab:systems}
  \begin{ruledtabular}
    \begin{tabular}{l c c c c r r}

      System &
      \multicolumn{1}{c}{\shortstack{Mismatch\\(\%)}} &
      \multicolumn{1}{c}{\shortstack{$\varepsilon_{\text{strain}}$\\(eV/atom)}} &
      \multicolumn{1}{c}{\shortstack{$S_{\text{Nb}}$\\(eV/\AA$^{2}$)}} &
      \multicolumn{1}{c}{\shortstack{$S_{\text{Ta}}$\\(eV/\AA$^{2}$)}} &
      \multicolumn{1}{c}{$N_{\text{ML}}^{\text{Nb}}$} &
      \multicolumn{1}{c}{$N_{\text{ML}}^{\text{Ta}}$} \\
      \hline
      AuPd      & -2.28 & 0.07 & -0.14 & -0.16 & 16 & 18 \\
      AuPt      & -2.11 & 0.10 & -0.17 & -0.19 & 13 & 15 \\
      AgPd      & -2.62 & 0.09 & -0.13 & -0.15 & 11 & 13 \\
      AgPt      & -2.51 & 0.14 & -0.17 & -0.19 &  9 & 10 \\
      Pd        & -4.90 & 0.13 & -0.08 & -0.12 &  5 &  7 \\
      Pt        & -4.03 & 0.14 & -0.07 & -0.07 &  4 &  4 \\
      Ag        &  0.05 & 0.03 & -0.08 & -0.09 & 20 & 22 \\ 
      Au        &  0.22 & 0.02 & -0.11 & -0.13 & 44 & 49 \\ \hline
      \multicolumn{7}{c}{Cu wetting-adhesion layer analysis} \\ \hline
      Au/Nb(3x1v)   & \multicolumn{1}{c}{--} & 0.03 & -0.04 &  0.10 & 11 &  0 \\
      Au/O/Nb      & \multicolumn{1}{c}{--} & 0.03 & -0.03 & -0.06 &  8 & 15 \\
      Au/Cu/Nb     & \multicolumn{1}{c}{--} & 0.03 & -0.30 & -0.25 & 78 & 66 \\
      Au/Cu/Nb(3x1v) & \multicolumn{1}{c}{--} & 0.03 & -0.32 & -0.11 & 84 & 27 \\
      Au/Cu/O/Nb   & \multicolumn{1}{c}{--} & 0.03 & -0.38 & -0.27 & 98 & 70 \\
    \end{tabular}
  \end{ruledtabular}
\end{table}

Fortunately, surface chemistry is controlled by the cap, not the substrate, once deposition exceeds a certain thickness. To determine this thickness, we calculated adsorption energies \(E_{\text{bind}}\) for O, N, and H on Nb(110) and Ta(110) covered by 1–3 ML of Au, Ag, Pd, or Pt (lowest-energy site, referenced to \(\tfrac12E_{X_2}\)).  In all cases \(E_{\text{bind}}\) increases (reflecting a reduced tendency to bind) relative to bare Nb or Ta.
Crucially, passivation \emph{saturates} by $\sim$2 ML, beyond which further reductions in binding are marginal (see End Matter). Thus, even an ultrathin layer, provided it wets, can chemically passivate the surface.

\ \\
\emph{Wetting/adhesion layers.---} Gold is a very promising candidate because it passivates well on ideal, oxygen-free Nb(110), requires extremely low strain, and, along with Ag, presents the lowest surface energy of the metals studied. However, real substrates contain steps and residual oxygen. Under these conditions Au tends to dewet, producing pinholes that admit oxygen. Indeed, recent experimental work by Wang \emph{et al.}~finds evidence of a tendency for three-dimensional growth of Au on oxygenated Nb surfaces~\cite{wang2025epitaxial}. Continuity is then restored only with a much thicker Au film, which exacerbates the proximity-induced suppression of $T_c$ and favors incoherent layer growth~\cite{oseroff2023measurements,de2024low,bal2024systematic}.

To decouple adhesion from passivation, we introduce the idea of using a wetting/adhesion underlayer (WAL) that bonds to both sides—Nb/Ta below and Au above—so the Au cap can remain within the 2–3\,ML passivation window while keeping the normal-metal thickness within the superconducting budget (see Fig.~\ref{fig:wal_diagram}).

\squeezetable 

\begin{table}
  \caption{Interface formation energies $\gamma_{\mathrm{int}}$ (eV/\AA$^2$) for candidate WAL metals on Nb and Ta substrates for three surface microstructures: clean (110) terrace, $3\times1$ vicinal surface, and O-decorated (110) surface. The metal/Au interface energy is also reported. More negative values indicate stronger adhesion.}
  \label{tab:interfaces-nb}
  \begin{ruledtabular}
    \begin{tabular}{l c c c c}
      System &
      \multicolumn{1}{c}{Clean interface} &
      \multicolumn{1}{c}{$3\times1$ interface} &
      \multicolumn{1}{c}{O decorated} &
      \multicolumn{1}{c}{Au interface} \\
      \hline
      Au/Nb & -0.03 &  0.04 &  0.05 &  0.00 \\
      Pd/Nb & -0.04 & -0.06 &  0.11 &  0.07 \\
      Pt/Nb & -0.06 & -0.12 &  0.13 &  0.10 \\
      Cu/Nb & -0.09 & -0.11 & -0.17 & -0.13 \\
      Zr/Nb &  0.03 &  0.10 & -0.06 &  0.04 \\
      Au/Ta & -0.02 &  0.21 &  0.05 &  0.00 \\
      Pd/Ta & -0.05 & -0.07 &  0.10 &  0.07 \\
      Pt/Ta & -0.02 & -0.09 &  0.13 &  0.10 \\
      Cu/Ta & -0.02 &  -0.00 & -0.17 & -0.13 \\
      Zr/Ta &  0.10 &  0.36 & -0.02 &  0.04 \\
    \end{tabular}
  \end{ruledtabular}
\end{table}

To explore this idea for both Nb(110) and Ta(110), we consider three representative microstructures as proxies for realistic imperfections:
(i) a clean terrace;
(ii) a $3\times1$ vicinal surface with single-atom steps, representative of oxide-driven surface reconstructions~\cite{mcmillan2020persistence}; and
(iii) an O-decorated surface with one monolayer of oxygen at hollow/bridge sites.
For each candidate WAL we compute the two formation energies relevant to the full
$\mathrm{substrate}/\mathrm{WAL}/\mathrm{Au}$ stack,
\[
\gamma_{\mathrm{int}}[\mathrm{sub}/\mathrm{WAL}]
\quad\text{and}\quad
\gamma_{\mathrm{int}}[\mathrm{WAL}/\mathrm{Au}].
\]
Table~\ref{tab:interfaces-nb} reports results for a representative subset of metals that capture the main trends (the full dataset is provided in the End Matter). We find the following:

\emph{Clean terraces (Nb and Ta).}
Candidate late-transition and noble metal WAL materials adhere reasonably on the substrate side of the sub/WAL/Au stack, but the binding of the WAL to Au on the other side of the junction separates candidates.
Cu exhibits the lowest $\gamma_{\mathrm{int}}[\mathrm{WAL}/\mathrm{Au}]$ while keeping $\gamma_{\mathrm{int}}[\mathrm{sub}/\mathrm{Cu}]$ small. This behavior aligns with prior wetting studies of Cu on Ta~\cite{dupraz2016wetting}.
Pd and Pt adhere well to Nb/Ta but pay a higher penalty against Au, placing them behind Cu as candidates, even on ideal terraces.

\emph{Vicinal $3\times1$ steps.}
Vicinal steps substantially increase $\gamma_{\mathrm{int}}$ for Au on both substrates (see the ``Au'' entries in Table~\ref{tab:interfaces-nb}). This underscores the need for a WAL whenever such imperfections are present. 
Pd and Pt reduce the substrate–side interface energy relative to Au, and Cu remains a strong all-around candidate.

\emph{Interfacial oxygen.}
An O monolayer strongly penalizes most $\mathrm{sub}/\mathrm{WAL}$ interfaces.
Cu is comparatively tolerant: $\gamma_{\mathrm{int}}[\mathrm{sub{+}O}/\mathrm{Cu}]$ is more favorable than for direct Au deposition on O-rich Nb/Ta.
By contrast, early transition metals such as Zr show negative $\gamma_{\mathrm{int}}$ on O-decorated Nb/Ta (favorable binding to oxygen) but positive values on clean/stepped surfaces and a higher cost against Au, identifying them as effective \emph{oxygen-getter} (sacrificial) layers rather than WALs. 

Finally, we note that while refractory metals such as Ru, Rh, and Ir also exhibit excellent passivation properties (high \(E_{\mathrm{inter}}\)), they possess significantly higher intrinsic surface energies (\(\gamma_{\mathrm{cap}}\)) compared to Au or Ag. Since wetting requires \(S =  \gamma_{\mathrm{int}} + \gamma_{\mathrm{cap}} - \gamma_{\mathrm{sub}} < 0\), this high surface energy imposes a severe thermodynamic penalty that is difficult to overcome even with an optimized interface. Consequently, while a WAL can improve the adhesion of refractory metals by lowering \(\gamma_{\mathrm{int}}\), Au remains the superior candidate because its low \(\gamma_{\mathrm{cap}}\) optimizes the spreading parameter.

Across substrates and microstructures, a 1–2\,ML Cu underlayer minimizes both interface energies and is the most robust WAL beneath Au. With adhesion secured by Cu, the Au termination can be limited to a few ML—enough to reach the weak-adsorption plateau for O/N/H—without exceeding the superconducting thickness budget (see Cu wetting-adhesion layer analysis section of Table \ref{tab:systems}).

\ \\
\emph{Impact of layers on superconductivity.---} Finally, to assess the impact of our proposed structures on substrate superconductivity, we now consider proximity effects in coupled superconducting--normal (S--N) bilayers with \(S = \mathrm{Nb}, \mathrm{Ta}\), using the strong-coupling Eliashberg formalism~\cite{ummarino2023strong, carbotte1990properties, pavarini2013emergent}.

Within this framework, the superconducting gaps \(\Delta_{j}(i\omega_n)\) and renormalization functions \(Z_{j}(i\omega_n)\) for a layer \(j \in \{S,N\}\) coupled to a neighboring layer \(k \neq j\) obey the coupled equations:

\begin{equation}
\begin{aligned}
Z_j(i \omega_n) \Delta_j(i \omega_n) &= \pi T \sum_m \left[\Lambda_j(i \omega_n, i \omega_m) - \mu_j^*(\omega_c)\right] \\
&\times \Theta(\omega_c - |\omega_m|) N_j^{\Delta}(i \omega_m) + \Gamma_j N_k^{\Delta}(i \omega_n)
\end{aligned}
\label{eq:gap}
\end{equation}

\begin{equation}
\omega_n Z_j(i \omega_n) = \omega_n + \pi T \sum_m \Lambda_j(i \omega_n, i \omega_m) N_j^Z(i \omega_m) + \Gamma_j N_k^Z(i \omega_n),
\label{eq:renorm}
\end{equation}
where  
\[
\Lambda_{S(N)}(i\omega_n,i\omega_m)=2\!\int_0^{\infty}\!d\Omega\;
\frac{\Omega\,\alpha^2_{S(N)}F(\Omega)}{(\omega_n-\omega_m)^2+\Omega^2},
\]
\(\omega_n\) are the Matsubara frequencies, \(\Theta\) is the Heaviside function, the \(\Gamma_{S(N)}\) encode interface transparency and thickness, and where $N_j^Z(i\omega_m)$ and $N_j^\Delta(i\omega_m)$ are the usual
dimensionless Eliashberg functions describing, respectively, the
quasiparticle and Cooper-pair densities of states in layer
$j \in \{S,N\}$, as defined in \cite{ummarino2023strong}.

Here, we work in the thin-film (Cooper) limit, where the superconducting and normal layers have
thicknesses \(d_S\) and \(d_N\) that are small compared to their respective coherence lengths
\(\xi_S\) and \(\xi_N\) (i.e., \(d_S\ll\xi_S\) and \(d_N\ll\xi_N\)). In this regime the order parameter
may be taken approximately uniform within each layer and the two films couple primarily through
the interface. This description is directly applicable to thin-film superconducting circuits and
ultrathin caps; for thick substrates (e.g., SRF cavities) we use it as a comparative trend model, interpreting \(d_S\) as an effective near-surface superconducting thickness participating in
the proximity coupling.

In the proximity-Eliashberg tunneling formulation of Ref.~\cite{ummarino2023strong}, the interface
coupling is encoded by effective rates $
\Gamma_{S(N)}=\pi |t|^{2}A\,d_{N(S)} N_{N(S)}(0),$
where \(A\) is the interface area, \(|t|^{2}\) an effective transmission factor, and \(N_{S,N}(0)\)
are the Fermi-level densities of states in S and N. Taking the ratio removes the unknown prefactors,
\[
\frac{\Gamma_S}{\Gamma_N}=\frac{d_N N_N(0)}{d_S N_S(0)}\approx \frac{d_N}{d_S},
\]
where the final approximation (common in the Cooper limit)
assumes \(N_N(0)\) and \(N_S(0)\) are comparable~\cite{ummarino2023strong}.

For our calculations, \(\alpha^2F(\Omega)\) was calculated \emph{ab initio} for all relevant layers. 
The resulting electron–phonon couplings are
\(\lambda_{\mathrm{Nb}}=0.95\),
\(\lambda_{\mathrm{Ta}}=0.94\),
\(\lambda_{\mathrm{Au}}=0.18\),
\(\lambda_{\mathrm{Pd}}=0.26\),
\(\lambda_{\mathrm{Pt}}=0.41\),
\(\lambda_{\mathrm{Ag}}=0.12\),
\(\lambda_{\mathrm{Cu}}=0.12\), and
\(\lambda_{\mathrm{Zr}}=0.70\),
all within representative experimental ranges~\cite{grimvall1976electron,allen1987empirical,tong2019comprehensive,papaconstantopoulos1977calculations,giri2020electron,chatterjee1980electron,al1987electron,bose2008electron}.  
We adopt $\mu^{*}_{\mathrm{Nb}}=0.14$~\cite{lee1996linear}, which gives $T_c^{\mathrm{Nb}}\approx 9.2$~K, in excellent agreement with experiment. For Ta, we use $\mu^{*}_{\mathrm{Ta}}=0.13$~\cite{al1987electron}, obtaining $T_c^{\mathrm{Ta}}\approx 7.5$~K; this overestimates the experimental value but is consistent with prior \textit{ab initio} reports~\cite{al1987electron}. For the normal-metal layers we adopt element-specific Coulomb pseudopotentials guided by Eliashberg analyses in the literature, taking \(\mu^{\!*}_{\mathrm{Au}}=\mu^{\!*}_{\mathrm{Ag}}=\mu^{\!*}_{\mathrm{Cu}}=0.11\)~\cite{ummarino2024can}, \(\mu^{\!*}_{\mathrm{Pd}}=0.12\)~\cite{papaconstantopoulos1978band}, \(\mu^{\!*}_{\mathrm{Pt}}=0.10\)~\cite{szczesniak2015high}, and \(\mu^{\!*}_{\mathrm{Zr}}=0.13\)~\cite{abe2018high}. All calculations employed \(\omega_c=120\)~meV and were solved self-consistently with a temperature spacing $T = T_c/20$ to reach numerical convergence.

\begin{figure}[htbp]
  \centering
  \includegraphics[width=\columnwidth]{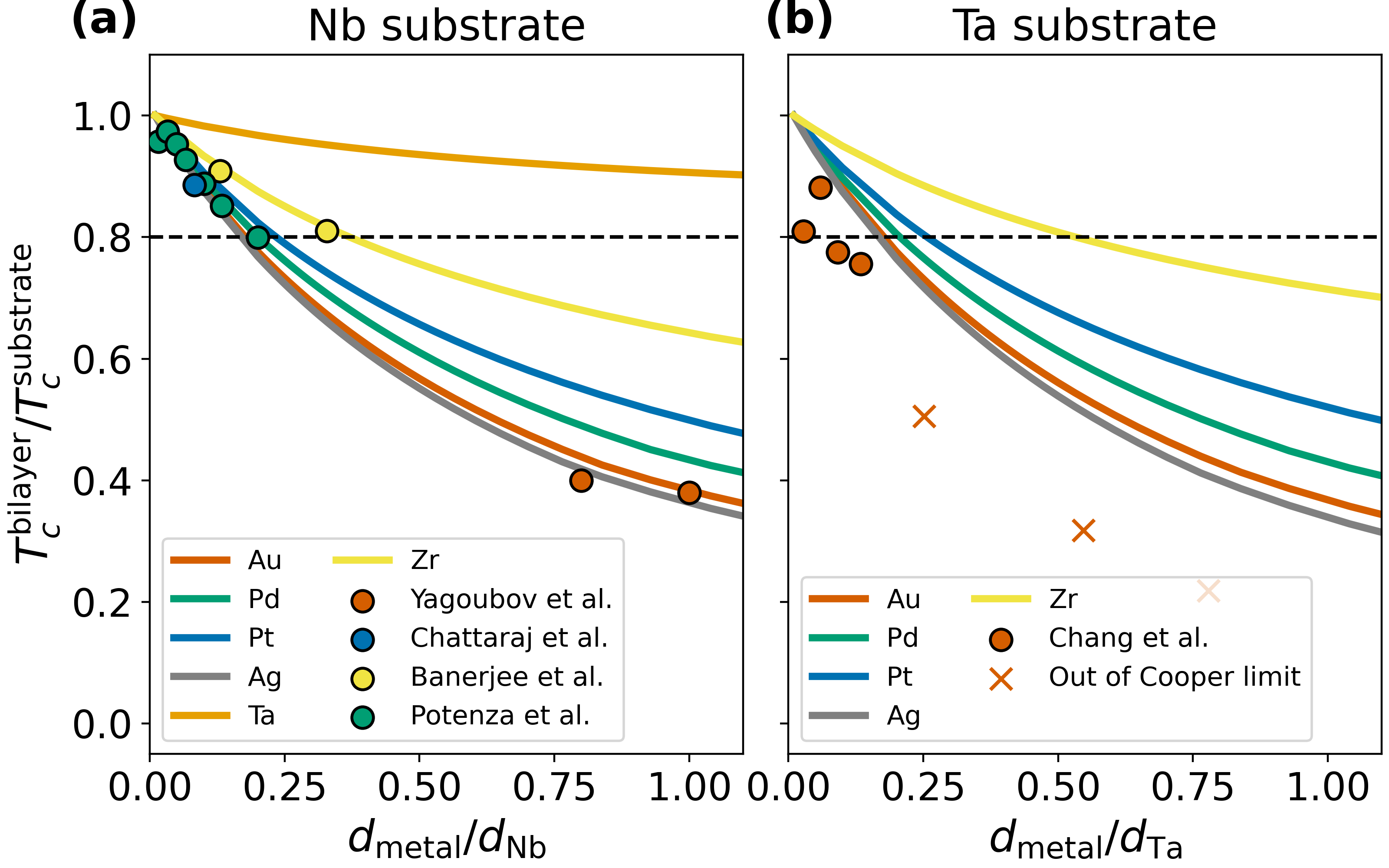}
  \caption{Normalized transition temperature for Nb\(|\)N (a) and
           Ta\(|\)N (b) bilayers versus \(d_N/d_S\).
            Horizontal black dashed line: \(T_c^{\text{bilayer}}/T_c^{\text{substrate}}\ge 0.8\). Experimental data are shown as solid dots \cite{yagoubov2001development,chattaraj2025preventing,banerjee2003proximity,potenza2007layer,chang2025eliminating}}.
  \label{fig:bilayer}
\end{figure}

Figure~\ref{fig:bilayer} shows the calculated
\(T_c/T_c^{\text{bulk}}\) for Nb/N and Ta/N bilayers versus the thickness ratio \(d_N/d_S\), together with available experimental data~\cite{yagoubov2001development,chattaraj2025preventing,banerjee2003proximity,potenza2007layer,chang2025eliminating}.
In general there is excellent agreement with our calculations in the Cooper limit, with the exception of Au on Ta substrates, which is thought to be due to the inverse-proximity effect recently studied in Ref.~\cite{lin2024inverse}.
In the small-thickness limit (\(d_N/d_S\sim0.1\)), an adhesion layer adds only a penalty of a few percent on either substrate, whereas increasing \(d_N/d_S\) rapidly suppresses \(T_c\).
The dashed band (\(T_c/T_c^{\text{bulk}}\ge 0.8\)) defines a conservative safe envelope.
Accordingly, the total normal-metal thickness should be kept in the low-\(d_N/d_S\) regime where adsorption already saturates; within this regime Au, Pd, and Pt perform comparably. Ta on Nb preserves $T_c/T_c^{\text{bulk}}\ge 0.9$ even for comparable thicknesses of the bilayers.
This arises from the improved proximity effect associated with the superconducting nature of Ta itself, explaining why Nb-based superconducting qubits capped with Ta perform better than substrates capped with Au at large thicknesses~\cite{bal2024systematic}. Zr is also attractive as a sacrificial getter: it preserves superconductivity relatively well and exhibits strongly favorable interstitial formation energies for attracting O, N, and H. We anticipate that the same design rule applies more broadly: for other superconducting substrates, capping layers that likewise preserve superconductivity as much as possible with increasing thickness should be preferred, especially when relatively thick overlayers are needed to prevent penetration of oxygen or other impurities.

\ \\
\emph{Conclusion.---}
By combining first–principles interfacial energetics, impurity thermodynamics, and strong-coupling Eliashberg theory, we have established a single quantitative framework for engineering metallic caps on Nb and Ta.  
Two potential roles emerge for ultrathin capping layers.
(i)  A passivation layer to reject O/N/H at surface level; Au, Ag, and their Pd/Pt-rich alloys (AuPd, AuPt, AgPd, AgPt) satisfy this requirement on both substrates, with Au being particularly favorable due to its low lattice mismatch and strain.
(ii)  A wetting/adhesion underlayer to bind strongly to the substrate and to the selected passivation layer (e.g., Au). Among the metals surveyed, a thin Cu film uniquely minimizes the two-sided interface energy even on realistic stepped or O-contaminated substrates.  
We also studied Zr as a sacrificial getter to trap impurities and replace the substrate's native oxide.

Moreover, Eliashberg calculations show that noble-metal caps degrade superconducting performance with increasing thickness. Thus, a few-monolayer Cu WAL coated with a few monolayers of Au can chemically protect the substrate without causing significant superconducting loss. Consequently, the chemically thick Au layers sometimes used to enforce continuity are unnecessary and likely detrimental once full coverage is secured at the monolayer level using a Cu WAL.  

A practical consideration for Cu wetting/adhesion layers is interdiffusion across Au/Cu/Nb(Ta) interfaces during elevated-temperature steps. In practice, this risk can often be reduced with low-temperature deposition and limiting post-deposition anneals, while keeping the WAL as thin as possible to achieve film continuity.

Our Eliashberg results also show that superconducting caps (e.g., Ta on Nb) avoid the normal-metal proximity penalty observed with Au on Nb, consistent with broader device-loss trends~\cite{oseroff2023measurements,de2024low,bal2024systematic}. We further find that Zr, which can act as an effective getter overlayer, has relatively little impact on surface superconductivity, making it a promising candidate as well.  

Finally, our results for the interface and surface energies of different metals on Nb and Ta explain the success of noble-metal alloys in eliminating surface oxides with thinner capping layers than conventionally used~\cite{chang2025eliminating}. These design rules point to new stacks, Au/Cu/(Nb,Ta) and AuPt/(Nb,Ta), that warrant experimental exploration for SRF cavities and next-generation qubits.  

Overall, the methodology presented here applies broadly and provides a predictive framework for optimizing future quantum and accelerator devices.

\emph{Acknowledgements. ---} This work was supported by the US National Science Foundation under award PHY-1549132, the Center for Bright Beams.

\emph{Data availability. ---}The data that support the findings of this article are openly available \cite{mendez2025_zenodo_oxide_caps}.

\clearpage
\bibliographystyle{myunsrt}
\bibliography{PRL}

\clearpage

\section*{End Matter}

\paragraph*{Sensitivity to functional choice and spin--orbit coupling.---}
All production calculations in the main text employ the PBEsol functional, chosen for its improved description of lattice parameters and surface energies in metals relative to standard GGA-PBE. To gauge how our key impurity energetics depend on this choice, we recomputed the O/N/H interstitial formation energies in bulk Nb using PBE. Relative to our PBEsol reference values
\[
(E_{\mathrm{inter}}^{\mathrm{O}}, E_{\mathrm{inter}}^{\mathrm{N}}, E_{\mathrm{inter}}^{\mathrm{H}})
= (-4.41,-2.62,-0.51)\,\mathrm{eV},
\]
the corresponding PBE values
\[
(E_{\mathrm{inter}}^{\mathrm{O}}, E_{\mathrm{inter}}^{\mathrm{N}}, E_{\mathrm{inter}}^{\mathrm{H}})
= (-4.37,-2.47,-0.44)\,\mathrm{eV}
\]
differ by at most $0.15$~eV. These shifts are comparable to the typical GGA-level uncertainty and are small compared to the multi-eV spread of \(E_{\mathrm{inter}}^{X}\) across the metals in Table~\ref{tab:interstitials}. In particular, they do not change the classification of Nb as a strong getter, nor the qualitative separation between early-transition-metal getters and late-transition-metal passivating layers. We therefore expect the design rules and periodic trends discussed in the main text to be robust with respect to the choice between PBE and PBEsol, and we adopt PBEsol for all survey calculations.

We also assessed the impact of spin--orbit coupling (SOC) for a representative heavy element. For the oxygen interstitial in Pt, the PBEsol formation energy increases from $E^{O}_{\mathrm{inter}} = 1.18$~eV in a scalar-relativistic calculation (no SOC) to $E^{O}_{\mathrm{inter}} = 1.48$~eV when using a fully relativistic pseudopotential including SOC, corresponding to a shift of approximately $0.3$~eV. While non-negligible in absolute terms, this shift is small compared to the overall spread of oxygen interstitial formation energies across the materials surveyed in this work, which range from approximately $-7.7$~eV to $+1.8$~eV (Table~\ref{tab:interstitials}). Importantly, the inclusion of SOC does not alter the qualitative ranking of materials or the classification into strong getter and passivation regimes relevant to the design rules discussed here. Because the central results of this study rely on relative energetic trends and large energy-scale separations rather than fine quantitative differences, we neglect SOC in the production calculations and base all reported design rules and periodic trends on scalar-relativistic PBEsol results, enabling consistent comparisons across the full material survey.

\paragraph*{Details of WAL.---} Table~\ref{tab:interfaces-nb-full} presents the detailed interface formation energies for the candidate WAL metals. While 25 elemental metals were initially surveyed (as listed in Table~\ref{tab:interstitials}), those exhibiting highly unfavorable clean interface energies on Nb (see Fig.~\ref{fig:islandMaps}) have been omitted from these tables, as they are unsuitable candidates for wetting or adhesion layers.

\begin{table}
  \caption{Interface formation energies $\gamma_{\mathrm{int}}$ (eV/\AA$^2$) for candidate WAL metals on Nb and Ta substrate for three surface microstructures: clean (110) terrace, $3\times1$ vicinal surface, and O-decorated (110) surface. The metal/Au interface energy is also reported. More negative values indicate stronger adhesion.}
  \label{tab:interfaces-nb-full}
  \begin{ruledtabular}
    \begin{tabular}{l c c c c}
      System &
      \multicolumn{1}{c}{Clean interface} &
      \multicolumn{1}{c}{$3\times1$ interface} &
      \multicolumn{1}{c}{O decorated} &
      \multicolumn{1}{c}{Au interface} \\
      \hline
      Pt/Nb & -0.06 & -0.12 &  0.12 &  0.10 \\
      Au/Nb & -0.03 &  0.04 &  0.05 &  0.00 \\
      Pd/Nb & -0.04 & -0.06 &  0.11 &  0.07 \\
      W/Nb  &  0.01 &  0.19 &  0.23 &  0.17 \\
      Cu/Nb & -0.09 & -0.11 & -0.17 & -0.13 \\
      Mo/Nb & -0.01 &  0.16 &  0.18 &  0.13 \\
      Cr/Nb &  0.04 & -0.09 &  0.18 &  0.13 \\
      Pb/Nb &  0.01 &  0.09 &  0.02 &  0.04 \\
      Cd/Nb &  0.02 &  0.10 &  0.01 &  0.08 \\
      Al/Nb & -0.01 & -0.01 &  0.08 &  0.05 \\
      Ta/Nb &  0.00 &  0.15 &  0.04 &  0.09 \\
      V/Nb  &  0.00 &  0.10 &  0.09 &  0.09 \\
      Mg/Nb &  0.01 &  0.12 & -0.06 &  0.04 \\
      Hf/Nb &  0.00 &  0.11 & -0.03 &  0.03 \\
      Zr/Nb &  0.03 &  0.09 & -0.06 &  0.04 \\
      Ti/Nb &  0.04 &  0.09 & -0.04 &  0.08 \\
      Y/Nb  &  0.03 &  0.12 & -0.10 &  0.02 \\
      Sc/Nb &  0.01 &  0.09 & -0.14 &  0.02 \\
      Pt/Ta & -0.02 & -0.09 &  0.13 &  0.10 \\
      Au/Ta & -0.02 &  0.21 &  0.05 &  0.00 \\
      Pd/Ta & -0.05 & -0.07 &  0.10 &  0.07 \\
      W/Ta  &  0.02 &  0.64 &  0.22 &  0.17 \\
      Cu/Ta & -0.02 &  0.00 & -0.17 & -0.13 \\
      Mo/Ta &  0.02 &  0.52 &  0.20 &  0.13 \\
      Cr/Ta &  0.01 &  0.35 &  0.20 &  0.13 \\
      Pb/Ta & -0.01 &  0.09 &  0.02 &  0.04 \\
      Cd/Ta &  0.03 &  0.15 &  0.01 &  0.08 \\
      Al/Ta &  0.01 &  0.08 &  0.07 &  0.05 \\
      V/Ta  &  0.14 &  0.42 &  0.12 &  0.09 \\
      Mg/Ta &  0.02 &  0.15 & -0.03 &  0.04 \\
      Hf/Ta &  0.06 &  0.40 &  0.02 &  0.03 \\
      Zr/Ta &  0.10 &  0.36 & -0.02 &  0.04 \\
      Ti/Ta &  0.05 &  0.33 & -0.01 &  0.08 \\
      Y/Ta  &  0.02 &  0.49 &  0.04 &  0.02 \\
      Sc/Ta &  0.03 &  0.40 & -0.07 &  0.02 \\
    \end{tabular}
  \end{ruledtabular}
\end{table}

\paragraph*{Bilayer superconductivity calculations.---}

\begin{figure}[!h]
   \centering
   \includegraphics*[width=0.65\columnwidth]{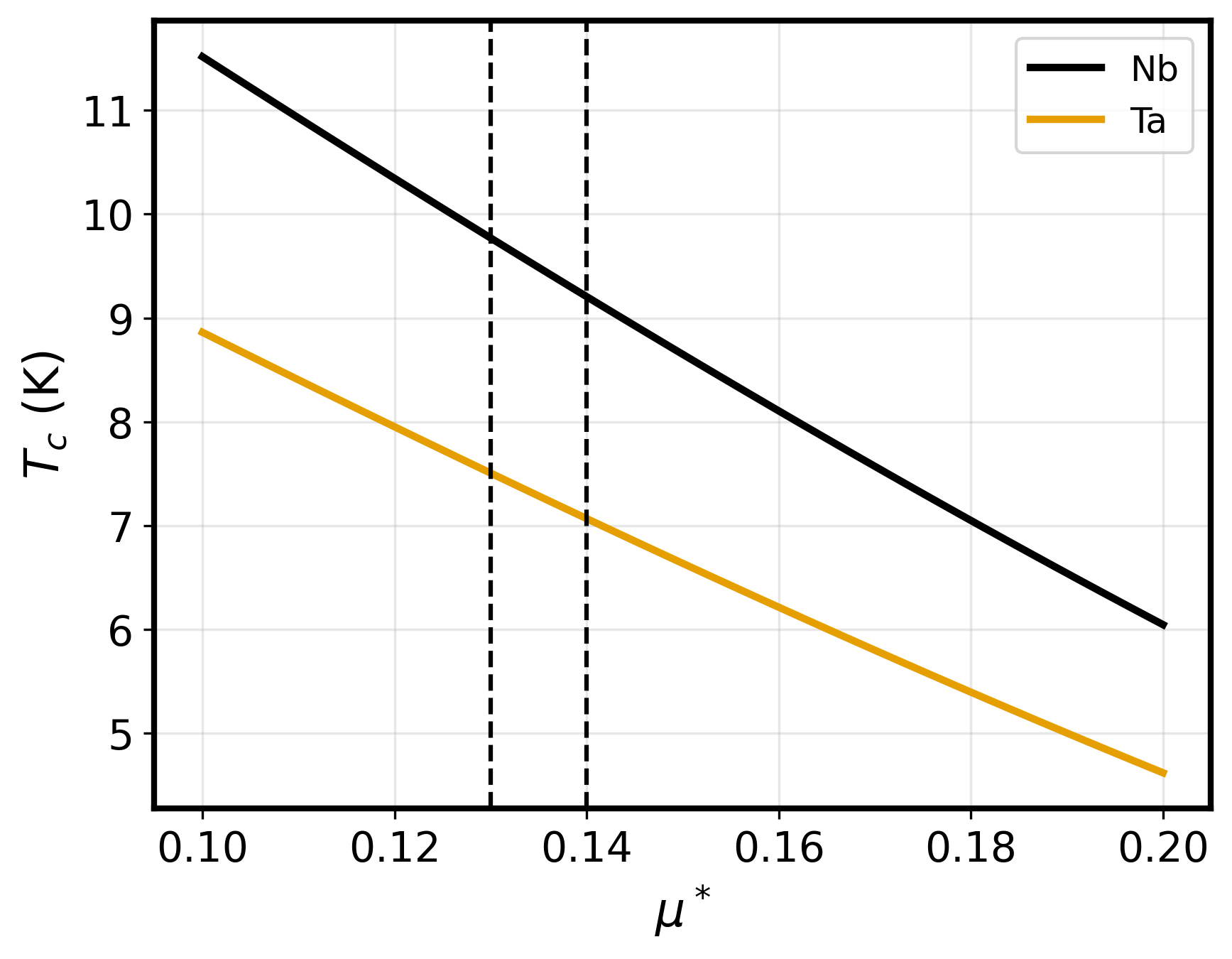}
   \caption{Calculated $T_c$ of Nb and Ta bulk as a function of $\mu^*$. The dashed lines represent $\mu^*=0.13$ and $\mu^*=0.14$ used in this letter for Ta and Nb respectively.}
   \label{fig:tc}
\end{figure}

\begin{figure}[!h]
   \centering
   \includegraphics*[width=0.7\columnwidth]{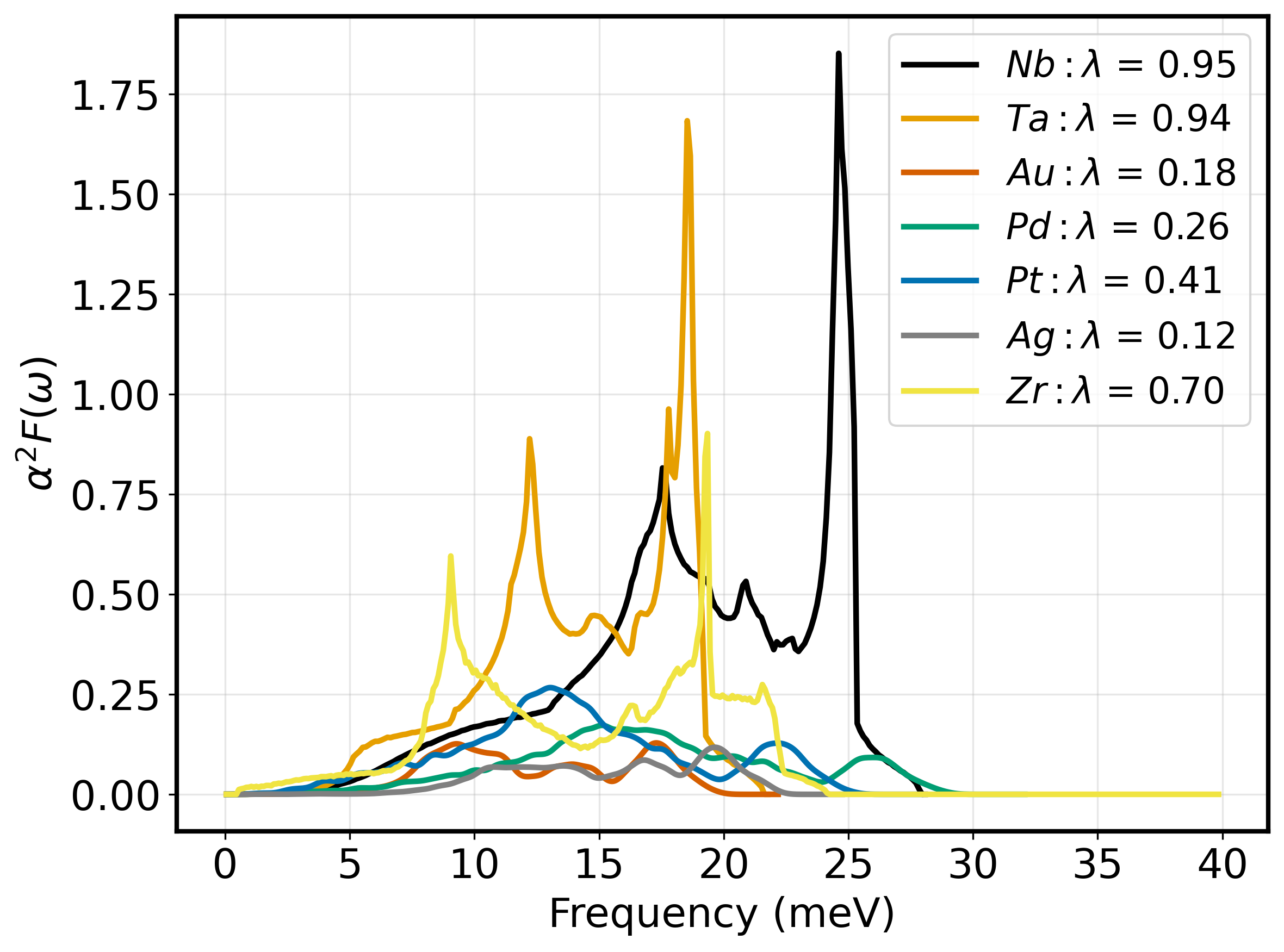}
   \caption{Computed Eliashberg spectral function for metals in our bilayer studies.}
   \label{fig:eliashberg}
\end{figure}

\begin{figure}[!h]
   \centering
   \includegraphics*[width=0.7\columnwidth]{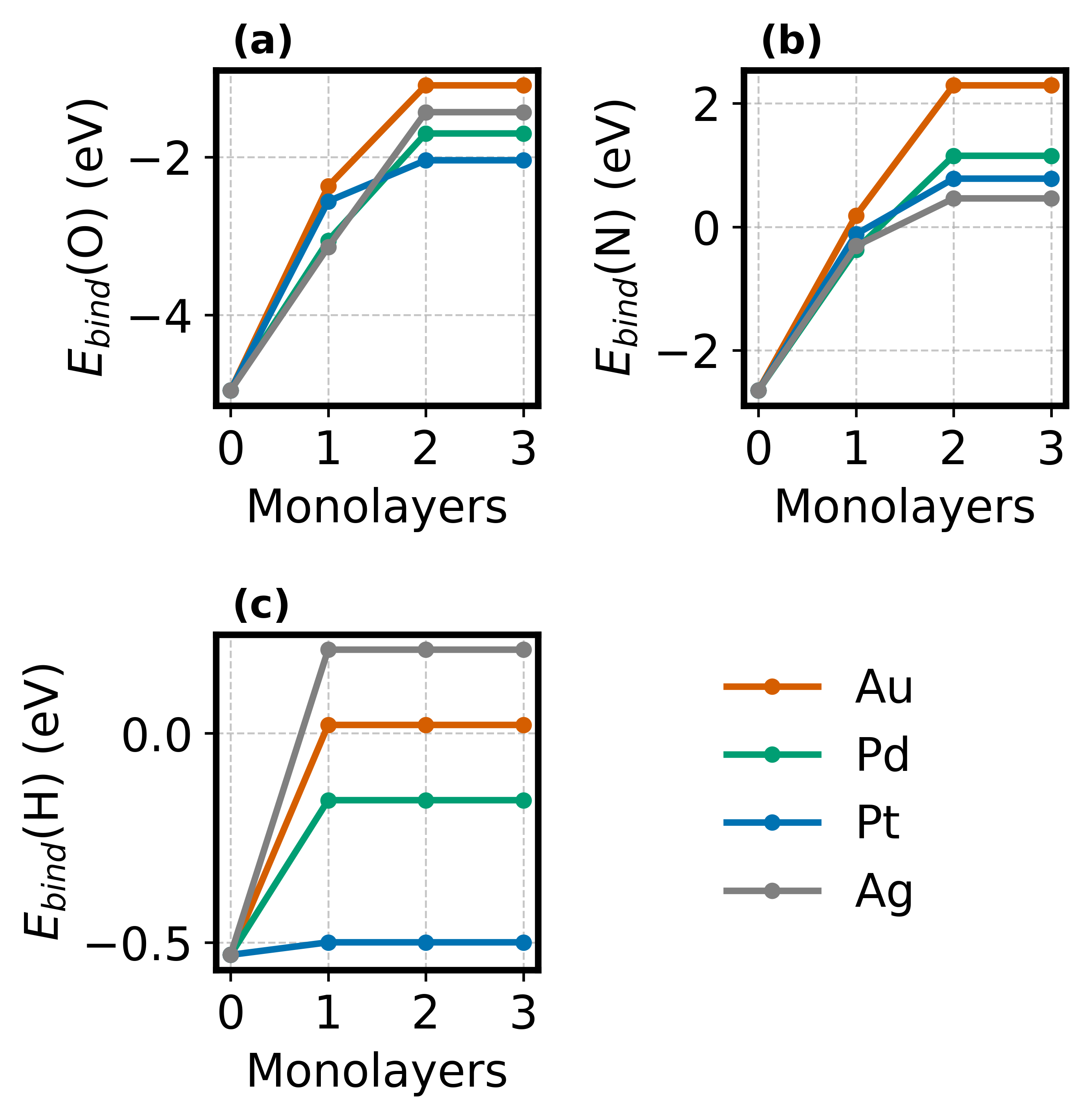}
   \caption{Nb(110) Binding energy to the surface ($E_{\text{bind}}>0$ is weaker adsorption) of (a) O, (b) N, and (c) H as a function
   of cap thickness for Au (orange), Pt (blue), Pd (green), and Ag (gray) on Nb(110).}
   \label{fig:bindNb}
\end{figure}

\begin{figure}[!h]
   \centering
   \includegraphics*[width=0.7\columnwidth]{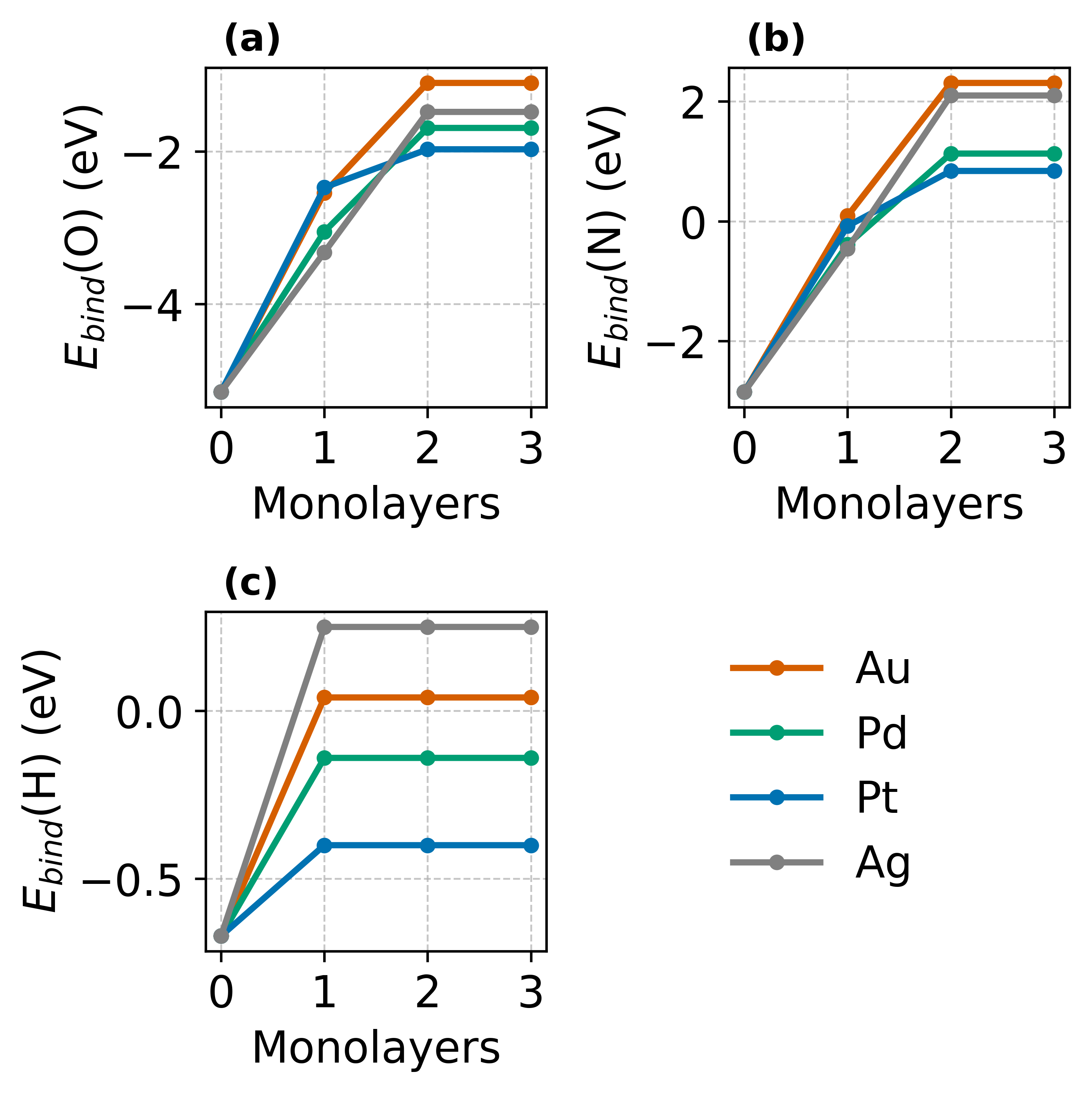}
   \caption{Ta(110) Binding energy to the surface ($E_{\text{bind}}>0$ is weaker adsorption) of (a) O, (b) N, and (c) H as a function
   of cap thickness for Au (orange), Pt (blue), Pd (green), and Ag (gray) on Ta(110).}
   \label{fig:bindTa}
\end{figure}

To illustrate the sensitivity of our results to the choice of the Coulomb pseudopotentials $\mu^*$, Fig.\ref{fig:tc} shows the corresponding values of the superconducting transition temperature $T_c$ for Nb and Ta.

Finally, Fig.~\ref{fig:eliashberg} shows the \emph{ab initio} Eliashberg spectral functions $\alpha^{2}F(\Omega)$ for all materials shown in Fig.~\ref{fig:bilayer}, from which we extract the corresponding electron--phonon coupling constants $\lambda$ reported in the text.

\paragraph*{Plots of surface impurity adsorption.---}
The main text referenced the convergence of impurity adsorption energies as a function of capping layer thickness. Figures~\ref{fig:bindNb} and \ref{fig:bindTa} display our results for O, N, and H, for both Nb and Ta capped with a variety of metals. The binding energy saturates within only $\sim$2 monolayers.

\end{document}